\documentclass[aps,prb,reprint,superscriptaddress]{revtex4-2}

\usepackage{amsmath, amssymb}
\usepackage{bm}
\usepackage{graphicx}
\usepackage{color}
\usepackage{dcolumn}
\usepackage{bm}
\usepackage[hidelinks]{hyperref}
\hypersetup{
  colorlinks   = true, 
  urlcolor     = blue, 
  linkcolor    = blue, 
  citecolor   = red 
}
\usepackage{multirow}
\usepackage{comment}

\allowdisplaybreaks[1]

\begin{document}

\title{
Coupled-cluster study of dynamic Jahn-Teller effect in a $5d^2$ W antifluorite 
}

\author{Teruki Matsuzaki}
\affiliation{Department of Chemistry, KU Leuven, Celestijnenlaan 200F, B-3001 Leuven, Belgium}
\affiliation{Graduate School of Science and Engineering, Chiba University, 1-33 Yayoi-cho, Inage-ku, Chiba-shi, Chiba 263-8522, Japan} 

\author{Liviu F. Chibotaru}
\email[]{liviu.chibotaru@kuleuven.be}
\affiliation{Theory of Nanomaterials Group, KU Leuven, Celestijnenlaan 200F, B-3001 Leuven, Belgium}

\author{Maristella Alessio}
\email[]{maristella.alessio@kuleuven.be}
\affiliation{Department of Chemistry, KU Leuven, Celestijnenlaan 200F, B-3001 Leuven, Belgium}

\author{Naoya Iwahara}
\email[]{naoya.iwahara@gmail.com}
\affiliation{Graduate School of Engineering, Chiba University, 1-33 Yayoi-cho, Inage-ku, Chiba-shi, Chiba 263-8522, Japan} 

\begin{abstract}
In correlated insulators, the interplay among coexisting charge, spin, orbital, and lattice degrees of freedom gives rise to rich quantum phenomena, while unraveling the interplay is not straightforward. 
In the family of cubic $5d^2$ double perovskites, the ground spin-orbit coupled electronic states of $5d$ metal sites are degenerate and couple to the Jahn-Teller active vibrations, whereas no experimental evidence of the symmetry-lowering in the low-temperature ordered phases has been reported.
To quantitatively unravel the nature of $5d^2$ centers, we apply equation-of-motion coupled cluster (EOM-CC) theory to analyze the vibronic and magnetic properties of $5d^2$ W sites of Cs$_2$WCl$_6$.
We derive the electronic and vibronic model Hamiltonians, calculate the W $L_3$ edge resonant inelastic x-ray scattering (RIXS) spectra, and determine the effective magnetic moment. 
The simulated RIXS spectra show that vibronic coupling makes several peaks asymmetric. 
The effective magnetic moments exhibit a temperature dependence similar to that observed experimentally, confirming the validity of the calculated distribution of low-energy levels. 
Our calculations indicate that the Jahn-Teller effect in Cs$_2$WCl$_6$ is in a weak regime, and noticeable deformation would not occur, whereas the dynamic Jahn-Teller effect modulates the shapes of the RIXS spectra and affects the magnetic moment. 
This work demonstrates the usefulness of the EOM-CC method for predicting physical phenomena on metal sites in correlated insulating materials. 
\end{abstract}

\maketitle

\section{Introduction}
\label{Sec_intro}
Heavy transition metal insulators show rich quantum magnetism \cite{Witczak-Krempa2014, Takagi2019, Takayama2021, Pourovskii2025}.
Multipolar orderings in $d^1$ \cite{Chen2010, Lu2017, Liu2018, Willa2019, Ishikawa2019, Hirai2020, Svoboda2021, Ishikawa2021Ta, Ishikawa2021Re, FioreMosca2021, MansouriTehrani2023, Kubo2023, Iwahara2023, Soh2024, FioreMosca2024, Muroi2025, Martinelli2026, Nikolov2026}, 
$d^2$ \cite{Chen2011, Maharaj2020, Paramekanti2020, Voleti2020, Lovesey2020, Pourovskii2021, Khaliullin2021, Takayama_RIKEN, Churchill2022, FioreMosca2022, Voleti2023, Rayyan2023, Pradhan2024, Omar2025, Hurt2025, Shibata2025}, 
and $d^3$ \cite{Pourovskii2023, Paddison2024} compounds, 
excitonic magnetism in $d^4$ compounds \cite{Khaliullin2013, Jain2017, Takahashi2021}, 
and Kitaev spin liquid phases and diverse magnetic ordered phases in $d^5$ compounds \cite{Jackeli2009, Rau2014, Yamaji2014, Wang2025}, to mention a few.
In heavy transition-metal-based insulating compounds, strong spin-orbit coupling at the metal sites is the primary source of the unconventional magnetism. 
A peculiar feature of heavy transition metal systems is the significant electron-phonon (vibronic) coupling arising from the strong covalency between the $d$ orbitals and those of neighboring ligands, which contrasts with other spin-orbit-coupled systems, such as $4f$ compounds.
Indeed, the interplay of the strong spin-orbit coupling and vibronic coupling in heavy transition metal compounds induces emergent phenomena: 
unconventional magnetic phases \cite{Liu2019, FioreMosca2021, Khaliullin2021, Iwahara2023, FioreMosca2024, Wang2025, Hurt2025, Benerjee2025, Martinelli2026, Mosca2026}, 
cross-correlated responses \cite{Iwahara2018, Tietjen2026, Banerjee2026}, 
generation of chiral phonon \cite{Sutcliffe2025},
and transport properties \cite{Minarro2026, Minarro2026b}.

\begin{figure}[tb]
\includegraphics[width=\linewidth, bb=0 0 921 449]{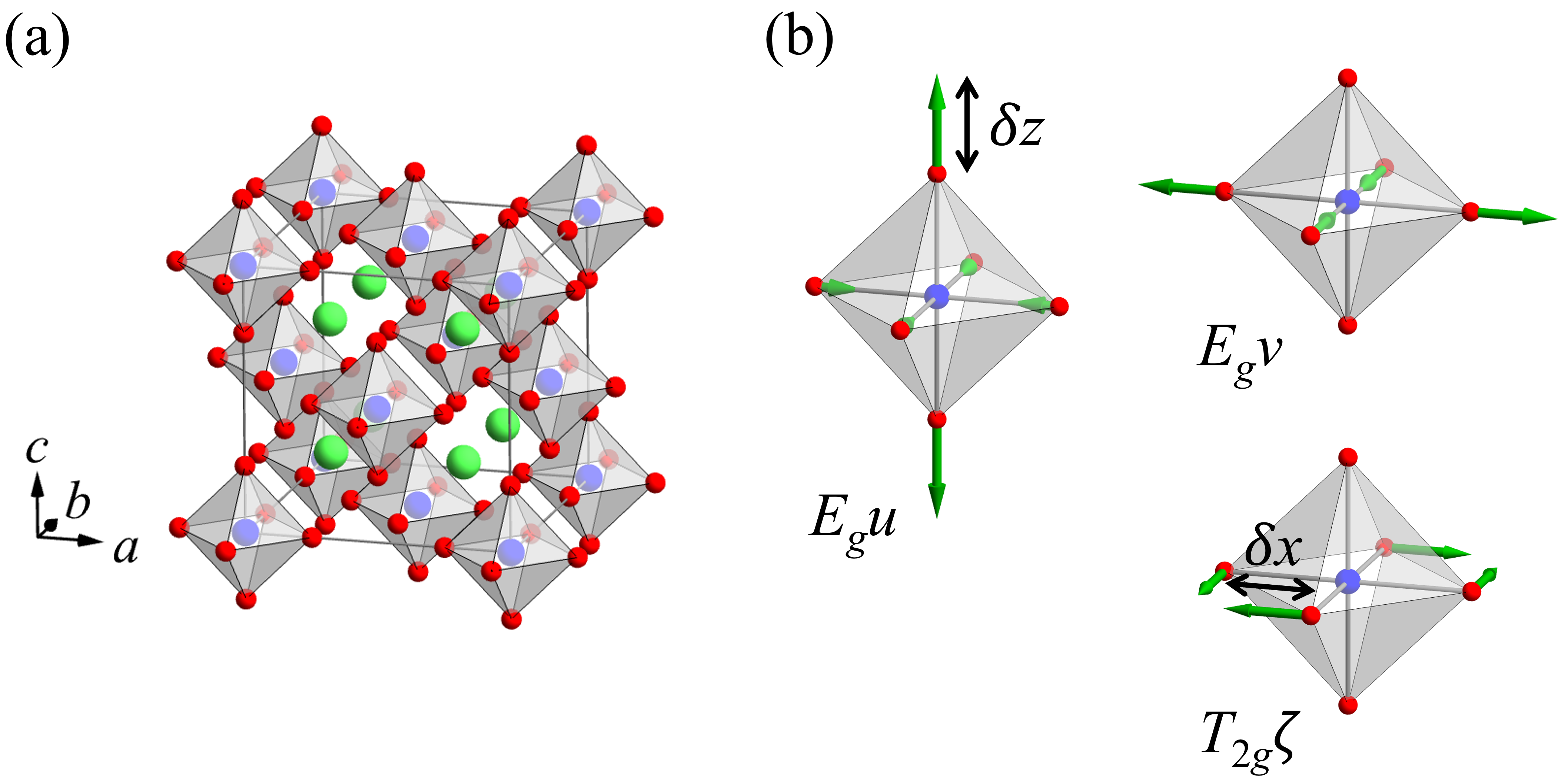}
\caption{(a) The conventional unit cell of antifluorite and (b) the JT active normal modes of the $5d^2$ octahedron.
The green, blue, and red spheres represent the Cs, W, and Cl atoms, respectively. 
(b) The green arrows indicate the displacements of the atoms by the JT active normal modes. 
}
\label{Fig_AF}
\end{figure}

The significance of the vibronic coupling is emphasized in the family of cubic $5d^1$ double perovskites. 
For $5d^1$ metal centers in these compounds, the ground electronic states are spin-orbit-coupled $J_\text{eff}=3/2$ quartet states \cite{Chen2010}. 
The $J_\text{eff}=3/2$ states couple to the Jahn-Teller (JT) active vibrations of the nearest neighbor ligand atoms \cite{Moffitt1957b, Englman1972, Iwahara2024}, which causes the structural transformations at low temperature 
in Ba$_2$NaOsO$_6$ \cite{Lu2017, Liu2018, Nikolov2026},
Ba$_2B$ReO$_6$ ($B=$ Mg, Ca) \cite{Hirai2020, Ishikawa2021Re,  Soh2024,  Muroi2025},
and Cs$_2$TaCl$_6$ \cite{Ishikawa2019, Ishikawa2021Ta, MansouriTehrani2023}.
The JT effect is a dynamic type \cite{Moffitt1957a, Moffitt1957b, Longuet-Higgins1958, Englman1972, Iwahara2024}, which has been confirmed by using resonant inelastic x-ray scattering (RIXS) measurements \cite{Frontini2024, Agrestini2024, Zivkovic2024, Iwahara2025, Matsuzaki2026}.
Furthermore, the theory based on the dynamic JT (DJT) effect on metal sites explains all the reported multipolar ordered phases in the family of $5d^1$ double perovskites \cite{Iwahara2023}.

On the contrary, the role of vibronic coupling is unclear in the family of cubic $5d^2$ double perovskite compounds [Fig. \ref{Fig_AF}(a)].
In these compounds, the ground states on the $5d^2$ metal centers are either nonmagnetic $E_g$ or magnetic $T_{2g}$ states arising from the ground $J_\text{eff}=2$ spin-orbit coupled multiplet states within the $t_{2g}^2$ configurations.
The $E_g$ states are the most stable in the family of Ba$_2B$OsO$_6$ ($B=$ Ca, Mg, Zn, Cd) and $A_2$WCl$_6$ ($A =$ Rb, Cs) [Fig. \ref{Fig_AF}(a)] \cite{Maharaj2020, Pradhan2024}, and the $T_{2g}$ states are the most stable in Ba$_2B$ReO$_6$ ($B=$ Sc, Y) \cite{Frontini2025, Omar2025}.
In both cases, the ground electronic states couple to the JT active modes [Fig. \ref{Fig_AF}(b)] \cite{Moffitt1957b, Englman1972, Iwahara2024}. 

Although the $5d^2$ sites are JT active, the structural data show quenched static JT deformations, which have been attributed to the development of an octupolar phase. 
Structural data show that all $5d^2$ osmium double perovskites are cubic down to 3.5 K with an accuracy of 0.1\% of the lattice constant \cite{Yamaura2006, Thompson2014, Marjerrison2016}.
In the cubic $5d^2$ double perovskites with the ground $E_g$ states on $5d$ sites, emergence of several ordered phases has been proposed: quadrupolar \cite{Khaliullin2021, Churchill2022}, ferro-octupolar \cite{Maharaj2020, Paramekanti2020, Voleti2020, Pourovskii2021, Rayyan2023, Pradhan2024, Hurt2025, Shibata2025}, and ferri-octupolar \cite{Lovesey2020} ordered phase. 
In the quadrupolar-ordered phases, static JT deformations can appear, whereas in the octupolar-ordered phases, static JT deformations are quenched. 
The latter is consistent with the absence of the JT deformations in the structural data. 

The dynamic JT effect can be relevant in the $5d^2$ compounds, as in the case of $5d^1$ compounds. 
A recent theoretical study suggests that the low-lying excitations in the ferro-octupolar ordered phases can be consistent with inelastic neutron diffraction data if the structure is symmetry-lowered \cite{Pourovskii2021}. 
The contradictory situation could be explained by the emergence of the dynamic JT effect as in the $5d^1$ systems. 
A recent oxygen $K$-edge RIXS spectra of Ba$_2$CaOsO$_6$ shows fine structures, which could come from the dynamic JT effect on $5d^2$ sites \cite{Okamoto2025}.

To unravel the vibronic effects in the $5d^2$ double perovskites, accurate knowledge of the vibronic coupling is desired. 
To this end, the state-of-the-art equation-of-motion coupled-cluster (EOM-CC) method \cite{Krylov2008, Sneskov2012, Bartlett2012} is an indispensable theoretical tool.
The EOM-CC method has been benchmarked for various transition metal complexes, providing accurate spin-state energy gaps and magnetic properties \cite{Orms2018, Pokhilko2019, Maristella2021, Maristella2023, Kahler2023, Alessio2024, Alessio2025}.
Recently, we have demonstrated the usefulness of the EOM-CC approach for describing the magnetic and spectroscopic properties of Mott insulators \cite{Eversdijk2025, Matsuzaki2026}:
the superexchange interactions between neighboring Cu ions in cuprates \cite{Eversdijk2025} and the vibronic coupling and Re $L_3$ edge RIXS spectra of $5d^1$ Ba$_2$MgReO$_6$ \cite{Matsuzaki2026}. 
These works demonstrate the applicability of EOM-CC to the study of magnetic energy levels and properties of quantum materials.

In this work, we use the EOM-CC method to unravel the roles of JT and vibronic effects in $5d^2$ W sites in Cs$_2$WCl$_6$. 
In the compound, each $5d^2$ W octahedron is well isolated from the others \cite{Takayama2026, Morgan2023} [Fig. \ref{Fig_AF}(a)], unlike in other $5d^2$ double perovskites, which validates the assumption that intersite interactions between W sites can be ignored when discussing the physical properties of a single W site. 
This situation is well-suited to the EOM-CC method. 
With this method, we determine the model Hamiltonian for $5d^2$ sites, comprising ligand-field, Hund's, spin-orbit, and vibronic couplings. 
We numerically diagonalize the model Hamiltonian and, using the obtained energy eigenstates, calculate the W $L_3$ edge RIXS spectra and the effective magnetic moment. 
The calculated RIXS spectra and the temperature dependence of the effective magnetic moment agree with the experimental data \cite{Takayama2026, Morgan2023}. 
In particular, the latter provides evidence of an accurate description of the low-lying states on $5d^2$ W sites.
This work illustrates the ability of the EOM-CC method to predict magnetic properties and spectroscopic data for correlated insulating materials.

\begin{figure*}[tb]
\includegraphics[width=0.95\linewidth, bb = 0 0 748 411]{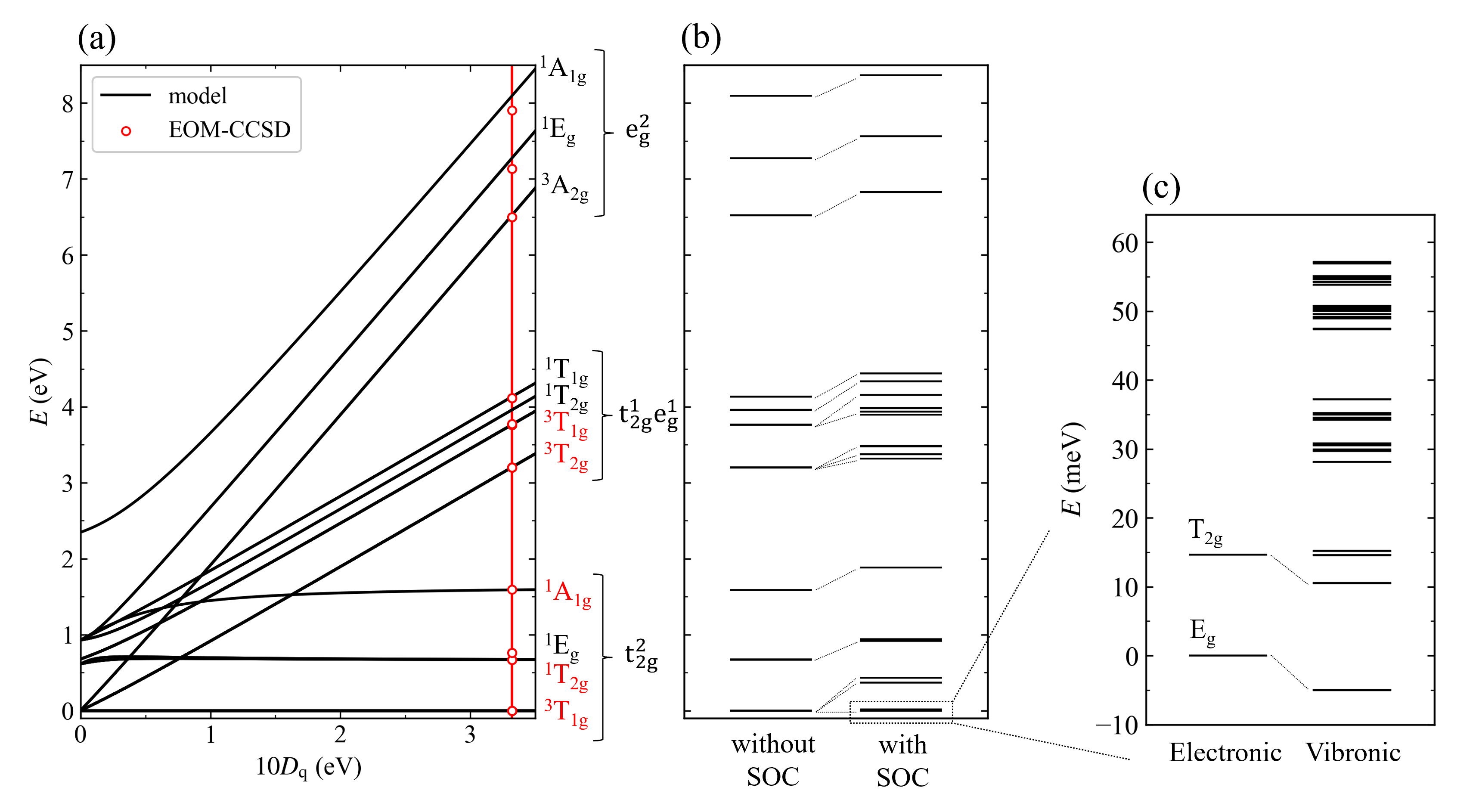}
\caption{
Energy levels of embedded $5d^2$ sites with respect to the ground energy level (in eV). 
(a) Tanabe-Sugano diagram. The black lines are the prediction by the Tanabe-Sugano diagram ($B=0.045$ eV, $C=0.193$ eV), and the red open circles and red solid lines indicate the {\it ab initio} levels and $10Dq$, respectively. 
(b) Comparison between the term levels (left) and spin-orbit multiplet states (right) obtained using the fitted data. We set the ground-state energy to zero.
(c) Low-lying $J_\text{eff}=2$ crystal-field levels (left) and the vibronic levels (right).}
\label{Fig_TS}
\end{figure*}

\section{Vibronic states of a $5d^2$ site in Cs$_2$WCl$_6$}
\label{Sec_model}
Figure \ref{Fig_AF}(a) shows the conventional cell of Cs$_2$WCl$_6$ antifluorite. 
In the system, [WCl$_6]^{2-}$ octahedra form an fcc lattice, and Cs$^+$ ions occupy tetrahedral sites. 
The antifluorite structure is a variant of the double perovskite structure with a vacancy at the $B$ site.

Cs$_2$WCl$_6$ is in a Mott insulating phase due to the large distance between $5d^2$ centers. 
In the insulating phase, the $5d$ electrons are localized on the W sites, and the nature of the $5d^2$ sites significantly influences the macroscopic physical phenomena. 
The local quantum states of an embedded $5d^2$ ion are determined by the interplay of the Hund's, spin-orbit, and vibronic couplings. 
On the embedded $5d^2$ ions, coexisting spin, orbital, and lattice degrees of freedom interact with one another via the ligand-field, Hund's coupling, spin-orbit coupling, and vibronic coupling, and form their quantum entanglement.
The ligand field and Hund's coupling on $d^2$ electrons cause the formation of term states, and spin-orbit coupling splits the term states into multiplet states \cite{Sugano1970}. 
The vibronic coupling between the degenerate multiplet states with the Jahn-Teller active vibrations [Fig. \ref{Fig_AF}(b)] gives rise to the spin-orbit-lattice entangled (vibronic) states \cite{Iwahara2024}.

Our electronic Hamiltonian for the $5d$ electrons consists of the atomic Coulomb interaction, ligand-field, and spin-orbit coupling \cite{Tanabe1954, Sugano1970}:
\begin{align}
 \hat{H}_\text{el} &= \hat{H}_\text{Coul} + \hat{H}_\text{SO} + \hat{H}_\text{LF}.
\end{align}
We employ the atomic Coulomb interaction between the $5d$ electrons:
\begin{align}
\hat{H}_\text{Coul} &= \frac{1}{2} \sum_{mnm'n'} \sum_{\sigma\sigma'} \langle mn\Vert m'n'\rangle \hat{d}_{m\sigma}^\dagger \hat{d}_{n\sigma'}^\dagger \hat{d}_{n'\sigma'} \hat{d}_{m'\sigma},
\end{align}
where $\hat{d}_{m\sigma}^\dagger$ and $\hat{d}_{m\sigma}$ are the creation and annihilation of an electron in the atomic $5d$ orbital with projection $m$ ($=-2,-1,0,+1,+2$) and spin $\sigma$ ($=-\frac{1}{2}, +\frac{1}{2}$), 
\begin{align}
\langle mn\Vert m'n'\rangle &= \delta_{m+n,m'+n'} (-1)^{m-m'} 
\sum_{k=0,2,4} F^k(dd)
\nonumber\\
&\times 
c^k(2m, 2m') c^k(2n,2n'),
\\
c^k(lm,l'm') &= \sqrt{\frac{[l']}{[l]}} (l0|l'0, k0) (lm|l'm',k m-m'),
\end{align}
$(lm|l'm',l''m'')$ are the Clebsch-Gordan coefficients for SO(3) group with Condon-Shortley's phase convention, 
and $F^k(dd)$ are the Slater-Koster parameters.
The Slater-Koster parameters are related to Racah's parameters as 
\begin{align}
 A = F^0 - \frac{F^4}{9}, 
 \quad
 B = \frac{1}{49} F^2 - \frac{5}{441} F^4, 
 \quad 
 C = \frac{5}{63} F^4. 
\end{align}
Below, we use Racah's parameters. 

The ligand-field Hamiltonian splits the $5d$ orbitals into the $e_g$ and $t_{2g}$ orbitals.
In terms of the spherical orbitals, the ligand-field (LF) Hamiltonian is 
\begin{align}
\hat{H}_\text{LF} &= 10Dq \sum_{mm'} \sum_\sigma \left(\bm{h}_\text{LF}\right)_{mm'} \hat{d}_{m\sigma}^\dagger \hat{d}_{m'\sigma},
\label{Eq_HLF}
\\
 \bm{h}_\text{LF} &=
 \begin{pmatrix}
  \frac{1}{10} & 0 & 0 & 0 & \frac{1}{2} \\
  0 & -\frac{2}{5} & 0 & 0 & 0 \\
  0 & 0 & \frac{3}{5} & 0 & 0 \\
  0 & 0 & 0 & -\frac{2}{5} & 0 \\
  \frac{1}{2} & 0 & 0 & 0 & \frac{1}{10} 
 \end{pmatrix}.
 \label{Eq:hLF}
\end{align}
Here, $10Dq$ is the ligand-field parameter, and the basis for $\bm{h}_\text{LF}$ is in the increasing order of the projection of orbital angular momentum.

In an atom, the spin-orbit coupling between the $d$ orbital and spin is 
\begin{align}
 \hat{H}_\text{SO} &= \lambda \sum_{m\sigma} \sum_{m'\sigma'} \langle m\sigma| \hat{\bm{l}} \cdot \hat{\bm{s}} |m'\sigma'\rangle \hat{d}_{m\sigma}^\dagger \hat{d}_{m'\sigma'},
 \label{Eq:HSO}
\end{align}
where $\lambda$ is the atomic spin-orbit coupling parameter, $\hat{\bm{l}}$ is the orbital angular momentum operator for the $5d$ orbitals, and $\hat{\bm{s}}$ is the spin angular momentum operator. 
This simple model for an isolated atom is convenient for understanding the concept, whereas the matrix elements of the Hamiltonian vary in crystals due to hybridization between the $d$ orbitals and neighboring ligand orbitals. 
We discuss the hybridization effect further in Sec. \ref{Sec_electronic}.

The ligand field on $5d$ levels and Hund's coupling between $5d$ electrons give rise to term states. 
According to the Tanabe-Sugano diagram \cite{Tanabe1954}, the ground term states are high-spin $t_{2g}^2$ $^3T_{1g}$ type [Fig. \ref{Fig_TS} (a)].

The spin-orbit coupling splits the $^3T_{1g}$ term levels into the spin-orbit multiplet states characterized by effective total angular momentum $J_\text{eff}=2$, 1, and 0 in the ascending order of energy. 
The five-fold degeneracy of the $J_\text{eff}=2$ multiplet states is weakly lifted by the Hund's and spin-orbit couplings between the $^3T_{1g}$ and excited term states involving the $e_g$ orbitals \cite{Voleti2020, Pradhan2024, Khaliullin2021}. 
The $J_\text{eff}=2$ splits into the $E_g$ and $T_{2g}$ multiplets; the former is lower in energy than the latter.

The $5d$ orbitals also couple to the JT active modes [Fig. \ref{Fig_AF}(b)].
The DJT model consists of the harmonic oscillator Hamiltonian $\hat{H}_0$ for the JT active modes and the vibronic coupling $\hat{H}_\text{JT}$, 
\begin{align}
 \hat{H}_\text{DJT} &= \hat{H}_0 + \hat{H}_\text{JT}, 
 \label{Eq:HDJT}
 \\
 \hat{H}_0 
 &=  \sum_{\Gamma = E_g, T_{2g}} \sum_{\gamma \in \Gamma} \frac{1}{2} \left( \hat{P}_{\Gamma\gamma}^2 + \omega_{\Gamma}^2 \hat{Q}_{\Gamma\gamma}^2 \right)
 \nonumber\\
 &=  \sum_{\Gamma = E_g, T_{2g}} \sum_{\gamma \in \Gamma} \hslash \omega_\Gamma \left( \hat{n}_{\Gamma\gamma} + \frac{1}{2} \right),
 \label{Eq_H0}
\end{align}
and 
\begin{widetext}
\begin{align}
 \hat{H}_\text{JT} &= 
 \sum_{\sigma} 
 \left(
  \hat{d}_{u \sigma}^\dagger,
  \hat{d}_{v \sigma}^\dagger,
  \hat{d}_{\xi \sigma}^\dagger,
  \hat{d}_{\eta \sigma}^\dagger,
  \hat{d}_{\zeta \sigma}^\dagger
  \right)
  \nonumber\\
  &\times
  \left[
  V_E
 \begin{pmatrix}
 -\hat{Q}_u   & \hat{Q}_v & 0 & 0 & 0 \\
  \hat{Q}_v & \hat{Q}_u   & 0 & 0 & 0 \\
  0 & 0 & -\frac{1}{2} \hat{Q}_u + \frac{\sqrt{3}}{2} \hat{Q}_v & 0 & 0 \\
  0 & 0 & 0 & -\frac{1}{2} \hat{Q}_u - \frac{\sqrt{3}}{2} \hat{Q}_v & 0 \\
  0 & 0 & 0 & 0 & \hat{Q}_u  \\
 \end{pmatrix}
 + 
 V_{T_2}
 \begin{pmatrix}
   0 & 0 & -\frac{1}{2} \hat{Q}_\xi & -\frac{1}{2} \hat{Q}_\eta & \hat{Q}_\zeta \\
   0 & 0 & \frac{\sqrt{3}}{2} \hat{Q}_\xi & -\frac{\sqrt{3}}{2} \hat{Q}_\eta & 0 \\
   -\frac{1}{2} \hat{Q}_\xi  &  \frac{\sqrt{3}}{2} \hat{Q}_\xi  & 0 & -\frac{\sqrt{3}}{2} \hat{Q}_\zeta & -\frac{\sqrt{3}}{2} \hat{Q}_\eta \\ 
   -\frac{1}{2} \hat{Q}_\eta & -\frac{\sqrt{3}}{2} \hat{Q}_\eta & -\frac{\sqrt{3}}{2} \hat{Q}_\zeta & 0 & -\frac{\sqrt{3}}{2} \hat{Q}_\xi \\
    \hat{Q}_\zeta            & 0                                & -\frac{\sqrt{3}}{2} \hat{Q}_\eta & -\frac{\sqrt{3}}{2} \hat{Q}_\xi & 0 \\
 \end{pmatrix}
 \right.
 \nonumber\\
  &+
 \left. 
  \frac{W_E}{2}
 \begin{pmatrix}
 -\{\hat{Q}_E^2\}_u   & \{\hat{Q}_E^2\}_v & 0 & 0 & 0 \\
  \{\hat{Q}_E^2\}_v & \{\hat{Q}_E^2\}_u   & 0 & 0 & 0 \\
  0 & 0 & -\frac{1}{2} \{\hat{Q}_E^2\}_u + \frac{\sqrt{3}}{2} \{\hat{Q}_E^2\}_v & 0 & 0 \\
  0 & 0 & 0 & -\frac{1}{2} \{\hat{Q}_E^2\}_u - \frac{\sqrt{3}}{2} \{\hat{Q}_E^2\}_v & 0 \\
  0 & 0 & 0 & 0 & \{\hat{Q}_E^2\}_u  \\
 \end{pmatrix}
 \right.
 \nonumber\\
 &+ 
 \left.
 \frac{W_E'}{2}
 \begin{pmatrix}
 -\{\hat{Q}_{T_2}^2\}_u   & \{\hat{Q}_{T_2}^2\}_v & 0 & 0 & 0 \\
  \{\hat{Q}_{T_2}^2\}_v & \{\hat{Q}_{T_2}^2\}_u   & 0 & 0 & 0 \\
  0 & 0 & -\frac{1}{2} \{\hat{Q}_{T_2}^2\}_u + \frac{\sqrt{3}}{2} \{\hat{Q}_{T_2}^2\}_v & 0 & 0 \\
  0 & 0 & 0 & -\frac{1}{2} \{\hat{Q}_{T_2}^2\}_u - \frac{\sqrt{3}}{2} \{\hat{Q}_{T_2}^2\}_v & 0 \\
  0 & 0 & 0 & 0 & \{\hat{Q}_{T_2}^2\}_u  \\
 \end{pmatrix}
 + \cdots
 \right]
 \begin{pmatrix}
  \hat{d}_{u \sigma} \\
  \hat{d}_{v \sigma} \\
  \hat{d}_{\xi \sigma} \\
  \hat{d}_{\eta \sigma} \\
  \hat{d}_{\zeta \sigma} \\
 \end{pmatrix},
 \label{Eq_HJT}
\end{align}
respectively.
Here, $u$, $v$, $\xi$, $\eta$, and $\zeta$ transform as $2z^2-x^2-y^2$, $x^2-y^2$, $yz$, $zx$, $xy$, under octahedral symmetry operations, respectively.  
$\hat{Q}_{\Gamma\gamma}$ and $\hat{P}_{\Gamma\gamma}$ are the mass-weighted normal coordinates and their conjugate momenta for the mode $\Gamma\gamma$ with frequency $\omega_{\Gamma}$. 
$\{\hat{Q}^2\}_{\Gamma\gamma}$ are the symmetrized normal coordinates (See, e.g., Ref. \cite{Matsuzaki2026}).
$V_\Gamma$ and $W_\Gamma$ are the linear and quadratic vibronic coupling parameters. 
The basis of the vibronic coupling \eqref{Eq_HJT} is the real $d$ orbital states.
The relation between the real and spherical harmonic $d$ orbital part is
\begin{align}
  \hat{d}_{u}^\dagger     &= \hat{d}_{0}^\dagger,
  \quad 
  \hat{d}_{v}^\dagger     = \frac{1}{\sqrt{2}} \left(\hat{d}_{-2}^\dagger + \hat{d}_{+2}^\dagger\right),
  \quad 
  \hat{d}_{\xi}^\dagger   = \frac{i}{\sqrt{2}} \left(\hat{d}_{-1}^\dagger + \hat{d}_{+1}^\dagger\right), 
  \quad 
  \hat{d}_{\eta}^\dagger  = \frac{1}{\sqrt{2}} \left(\hat{d}_{-1}^\dagger - \hat{d}_{+1}^\dagger\right), 
  \quad 
  \hat{d}_{\zeta}^\dagger = \frac{i}{\sqrt{2}} \left(\hat{d}_{-2}^\dagger - \hat{d}_{+2}^\dagger\right).
\end{align}
Here, $\sigma$ is omitted for simplicity. 

The spin-orbit coupling does not quench the vibronic coupling in the octahedral environment. 
The vibronic coupling between the ground ${}^3T_{1g}$ term states from $t_{2g}^2$ configurations and the JT active modes is 
\begin{align}
 \bm{H}_\text{JT}^{^3T_{1g}} &= 
 -V_E 
 \begin{pmatrix}
  -\frac{1}{2} \hat{Q}_u + \frac{\sqrt{3}}{2} \hat{Q}_v & 0 & 0 \\
  0 & -\frac{1}{2} \hat{Q}_u - \frac{\sqrt{3}}{2} \hat{Q}_v & 0 \\
  0 & 0 & \hat{Q}_u  \\
  \end{pmatrix}
  - V_T
  \begin{pmatrix}
    0 & -\frac{\sqrt{3}}{2} \hat{Q}_\zeta & -\frac{\sqrt{3}}{2} \hat{Q}_\eta \\ 
    -\frac{\sqrt{3}}{2} \hat{Q}_\zeta & 0 & -\frac{\sqrt{3}}{2} \hat{Q}_\xi \\
    -\frac{\sqrt{3}}{2} \hat{Q}_\eta & -\frac{\sqrt{3}}{2} \hat{Q}_\xi & 0 \\
  \end{pmatrix},
  \label{Eq_VJT_3T1g}
\end{align}
with the basis of $|{}^3T_{1g}\xi\rangle$, $|{}^3T_{1g}\eta\rangle$, and $|{}^3T_{1g}\zeta\rangle$. 
The spin part is not explicitly shown. 
Then, using the weakly split $J_\text{eff}=2$ multiplet states from the $^3T_{1g}$ term states as the basis, the linear vibronic Hamiltonian reduces to 
\begin{align}
 \bm{H}_\text{JT}^{J=2} &= \frac{V_E}{2}
 \begin{pmatrix}
 -\hat{Q}_u &  \hat{Q}_v & 0 & 0 & 0 \\
  \hat{Q}_v &  \hat{Q}_u   & 0 & 0 & 0 \\
  0 & 0 & -\frac{1}{2} \hat{Q}_u + \frac{\sqrt{3}}{2} \hat{Q}_v & 0 & 0 \\
  0 & 0 & 0 & -\frac{1}{2} \hat{Q}_u - \frac{\sqrt{3}}{2} \hat{Q}_v & 0 \\
  0 & 0 & 0 & 0 & \hat{Q}_u  \\
 \end{pmatrix}
 + 
 \frac{V_T}{2}  
 \begin{pmatrix}
  0 & 0 & -\frac{1}{2} \hat{Q}_\xi & -\frac{1}{2} \hat{Q}_\eta & \hat{Q}_\zeta \\
  0 & 0 & \frac{\sqrt{3}}{2}\hat{Q}_\xi & -\frac{\sqrt{3}}{2}\hat{Q}_\eta & 0 \\
  -\frac{1}{2} \hat{Q}_\xi  & \frac{\sqrt{3}}{2}\hat{Q}_\xi & 0 & -\frac{\sqrt{3}}{2} \hat{Q}_\zeta & -\frac{\sqrt{3}}{2} \hat{Q}_\eta \\
  -\frac{1}{2} \hat{Q}_\eta & -\frac{\sqrt{3}}{2}\hat{Q}_\eta & -\frac{\sqrt{3}}{2} \hat{Q}_\zeta & 0 & -\frac{\sqrt{3}}{2} \hat{Q}_\xi \\
  \hat{Q}_\zeta           &                              0 & -\frac{\sqrt{3}}{2} \hat{Q}_\eta  & -\frac{\sqrt{3}}{2} \hat{Q}_\xi & 0 
 \end{pmatrix}.
 \label{Eq_VJT_J}
\end{align}
\end{widetext}
The basis of $\bm{H}_\text{JT}^{J=2}$ is in the order of 
$|E_gu\rangle$,  $|E_gv\rangle$,  $|T_{2g}\xi\rangle$,  $|T_{2g}\eta\rangle$, and  $|T_{2g}\zeta\rangle$ from the ground $J_\text{eff}=2$ multiplet.
Within the above treatment, ignoring the hybridization between different term states, the spin-orbit coupling reduces the magnitude of the vibronic coupling by half compared with Eqs. \eqref{Eq_HJT} and \eqref{Eq_VJT_3T1g}.
We obtain similar forms for the quadratic vibronic couplings by the same transformations of the electronic states.

Vibronic coupling induces quantum entanglement between the electronic and lattice vibrational degrees of freedom. 
The vibronic coupling between the weakly split $J_\text{eff}=2$ states and the $E_g$ modes becomes $E \otimes E$ JT type in the $E_g$ state and $T_2 \otimes E$ JT type in the $T_{2g}$ state [see the $E_g$ vibronic term in Eq. \eqref{Eq_VJT_J}] \cite{Englman1972}. 
Since the quantum effect in the $E \otimes E$ JT system is stronger than that in the $T_2 \otimes E$ JT system, the vibronic coupling stabilizes the ground $E_g$ states more than the $T_{2g}$ states. 
The vibronic coupling to the $T_{2g}$ modes is in the first order in the $T_{2g}$ electronic states, while it does not exist within the $E_g$ states.  
Therefore, the vibronic coupling to the $T_{2g}$ modes stabilizes only the $T_{2g}$ states.
The vibronic coupling to the $E_g$ modes tends to be by a few times stronger than that to the $T_{2g}$ modes, and thus, the $E_g$ vibronic states are the most stable. 

We express the orbital-lattice entangled vibronic states on W centers as follows
\cite{Moffitt1957a, Moffitt1957b, Longuet-Higgins1958}:
\begin{align}
 |\Psi_\nu\rangle &= \sum_{i} |\Phi_i\rangle \otimes |\chi_{i;\nu}\rangle.
 \label{Eq_Psi}
\end{align}
Here, $|\Psi\rangle$ are the vibronic states, $|\Phi\rangle$ the eigenstates of $\hat{H}_\text{el}$, and $|\chi\rangle$ the lattice vibrational states. 
By using the eigenstates of the harmonic oscillator Hamiltonian \eqref{Eq_H0} as the basis for the lattice part, we expand $|\chi\rangle$ as 
\begin{align}
 |\chi_{i;\nu}\rangle &= \sum_{n_u n_v n_\xi n_\eta n_\zeta=0}^\infty |n_u n_v n_\xi n_\eta n_\zeta\rangle \chi_{i\bm{n}; \nu},
 \label{Eq_chi_n}
\end{align}
with $\bm{n} = (n_u, n_v, n_\xi, n_\eta, n_\zeta)$, and $\chi$ are the coefficients.
This representation of $|\chi\rangle$ is convenient to describe the vibronic states for weak to intermediate strength of the vibronic coupling \cite{Moffitt1957a, Moffitt1957b, Longuet-Higgins1958}. 

Before turning to the description of our {\it ab initio} approach, we briefly introduce the dimensionless vibronic coupling and coordinates. 
We transform the mass-weighted normal coordinates and the conjugate momenta into the dimensionless coordinates and momenta, respectively, by the following relations:
\begin{align}
 \hat{Q}_{\Gamma\gamma} = \sqrt{\frac{\hslash}{2\omega_{\Gamma}}} \hat{q}_{\Gamma\gamma},
 \quad 
 \hat{P}_{\Gamma\gamma} = \sqrt{\frac{\hslash\omega_{\Gamma}}{2}} \hat{p}_{\Gamma\gamma}.
\end{align}
With the dimensionless coordinates and momenta, we use the dimensionless vibronic couplings defined by 
\begin{align}
 g_{\Gamma} = \frac{V_{\Gamma}}{\sqrt{\hslash \omega_\Gamma^3}}, \quad 
 w_{E} = \frac{W_{E}}{\omega_E^2}, 
 \quad 
 w'_{E} = \frac{W'_{E}}{\omega_{T_2}^2}.
\end{align}
See Appendix A in Ref. \cite{Matsuzaki2026} for details.

\section{Computational details}
\label{Sec_comput}

In this section, we present the computational details used to derive the electronic Hamiltonian, vibronic coupling, vibronic states, RIXS spectra, and magnetic susceptibility. 

\subsection{Electronic states}
\label{Sec_electronic}

\begin{figure}[tb]
\includegraphics[width=0.8\linewidth, bb=0 0 494 475]{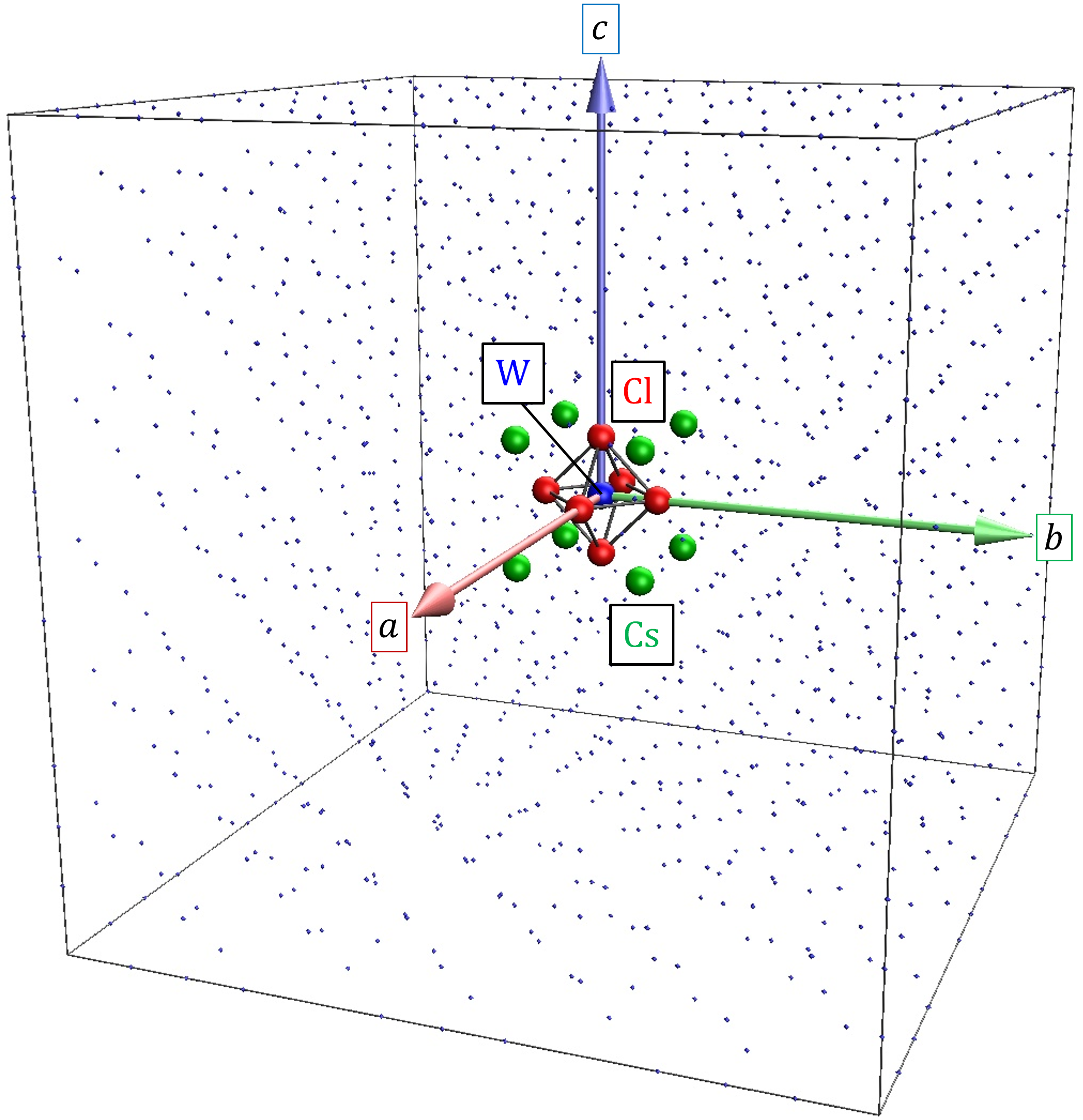}
\caption{
The tungsten cluster (WCl$_6$ and the nearest eight Ba ions) is embedded in point charges. The quantum-mechanical cluster is cut out from the crystallographic structure \cite{Morgan2023}. 
The blue dots represent the point charges. 
}
\label{Fig_cluster}
\end{figure}

We calculated the electronic states of a cluster of Cs$_2$WCl$_6$ using the equation-of-motion coupled-cluster singles and doubles (EOM-CCSD) method \cite{Krylov2008, Sneskov2012, Bartlett2012}. 
We performed electronic state calculations using the Q-Chem package (version 6.1, 6.3) \cite{Qchem} with basis sets taken from Basis Set Exchange \cite{Pritchard2019}.
 
The cluster is extracted from the experimental crystal structures \cite{Morgan2023}. 
The cluster consists of a central WCl$_6$ octahedron and its nearest-neighbor diamagnetic ions (eight Cs). 
We treat the Cs$_8$WCl$_6$ cluster quantum chemically. 
For the cluster, scalar relativistic effects are included via spin-free exact two-component theory in its one-electron variant (SFX2C-1e) \cite{Bolvin2017, Ilias2007, Liu2009}.
For the atoms of the cluster, we employed contracted basis sets as follows:
for the W ion, an all-electron type of the exact two-component triple zeta valence and polarization (X2C-TZVP) \cite{Pollak2017}, for the nearest neighbor six chlorine ligand atoms, the correlation-consistent polarized valence triple zeta (cc-pVTZ) basis set \cite{Woon1993}, and, for the next nearest neighbor Cs ions, Stevens, Basch, Krauss, Jasien, and Cundari's valence double zeta (SBKJC-VDZ) type with the corresponding effective core potentials for 54 core electrons of each Cs ion \cite{Stevens1992}.

To obtain the $5d^2$ term states of the W cluster, we employed the double-electron-attachment (DEA) variant of the EOM-CCSD approach \cite{Nooijen1995, Gulania2021}, in which two extra electrons are attached to a $5d^0$ reference state. 
In the EOM-CCSD calculations, 57 core electrons were frozen.
We computed eleven term states: $^3T_{1g}$, $^1T_{2g}$, $^1E_g$, $^1A_{1g}$, $^3T_{2g}$, $^3T_{1g}$, $^1T_{2g}$, $^1T_{1g}$, $^3A_{2g}$, $^1E_g$, and $^1A_{1g}$ levels in the increasing order of energy. 

To construct the electronic model, we used the EOM-CCSD data as follows. 
We derived the ligand-field and Racah parameters by fitting the calculated EOM-CCSD levels to an electronic Hamiltonian consisting of $\hat{H}_\text{LF}$ and $\hat{H}_\text{Coul}$.
For the fitting, we used the low-energy $^3T_{1g}$, $^1T_{2g}$, $^1A_{1g}$, $^3T_{2g}$, $^3T_{1g}$ term levels. 

Using the EOM-CCSD wave functions, we calculated the Breit-Pauli spin-orbit Hamiltonian matrix using the implemented method in the Q-Chem package \cite{Pokhilko2019, Pokhilko2019b, Carreras2020}.
Since the degree of $5d$-$3p$ hybridization between the tungsten $5d$ orbitals and ligand $3p$ orbitals, and hence, the matrix elements of the electronic operators depend on the terms, the spin-orbit coupling becomes as follows:
\begin{align}
 \hat{H}_\text{SO} &= 
 \sum_{S\Gamma M_S\gamma} 
 \sum_{S'\Gamma' M_S'\gamma'} 
 \lambda_{^{[S]}\Gamma\text{-}^{[S']}\Gamma'}
 \Bigg(
 \sum_{m\sigma} 
 \sum_{m'\sigma'} 
 \langle m\sigma| \hat{\bm{l}}\cdot \hat{\bm{s}} |m'\sigma'\rangle 
 \nonumber\\
 &\times 
 \langle {}^{[S]}\Gamma M_S\gamma|\hat{d}_{m\sigma}^\dagger \hat{d}_{m'\sigma'}|{}^{[S']}\Gamma' M_S'\gamma'\rangle 
 \Bigg)
 \nonumber\\
 &\times 
 |{}^{[S]}\Gamma M_S\gamma\rangle \langle{}^{[S']}\Gamma' M_S'\gamma'|,
 \label{Eq_HSO_cov}
\end{align}
where $|{}^{[S]}\Gamma M_S\gamma\rangle$ are the term states. 
We derived the spin-orbit coupling parameters $\lambda_{^{[S]}\Gamma\text{-}^{[S']}\Gamma'}$ by fitting the {\it ab initio} spin-orbit matrix elements to the model, Eq. \eqref{Eq_HSO_cov}.
Since the spin-orbit coupling dominantly comes from the $5d$ orbitals of the W ion, the covalency tends to reduce the spin-orbit coupling parameters compared with the spin-orbit coupling parameter for an atom. 

We derived the orbital angular momenta in a similar manner to the spin-orbit coupling. 
We included the $5d$-ligand $ 3p$-hybridization effect on the orbital angular momenta of the $t_{2g}$ orbitals via reduction factors. 
The orbital angular momenta are 
\begin{align}
 \hat{\bm{L}} &= 
 \sum_{S\Gamma M_S\gamma} 
 \sum_{\Gamma' \gamma'} 
 k_{^{[S]}\Gamma\text{-}^{[S]}\Gamma'} 
 \nonumber\\
 &\times
 \left(
 \sum_{mm' \in d} \sum_\sigma \langle m| \hat{\bm{l}} |m'\rangle 
 \langle {}^{[S]}\Gamma \gamma|\hat{d}_{m\sigma}^\dagger \hat{d}_{m'\sigma}|{}^{[S]}\Gamma' \gamma'\rangle 
 \right)
 \nonumber\\
 &\times
 |{}^{[S]}\Gamma M_S\gamma\rangle \langle{}^{[S]}\Gamma' M_S\gamma'|.
 \label{Eq_L_cov}
\end{align}
Here $k_{^{[S]}\Gamma\text{-}^{[S]}\Gamma'}$ are the reduction factors with respect to the atomic value. 
We obtained the reduction factors by fitting the EOM-CC orbital angular momenta matrix to Eq. \eqref{Eq_L_cov}.

\subsection{Vibronic coupling}
\label{Sec_vibronic}
We calculated the vibronic coupling parameters for the $t_{2g}$ orbitals from the gradient of the EOM-CCSD adiabatic potential energy surfaces (APES) of the ground $^3T_{1g}$ term. 
We generated several tens of deformed structures by using the relation between the mass-weighted normal coordinates for the JT active modes and the Cartesian coordinates \cite{Wilson1980}:
\begin{align}
 \bm{R}_A &= \bm{R}_A^{0} + Q_\alpha \frac{\bm{u}^\alpha_A}{\sqrt{M_A}},
\end{align}
where $\bm{u}^\alpha$ are the eigenvectors of the dynamical matrix, $M_A$ the mass of atom $A$, $Q_\alpha$ the mass-weighted normal coordinates, and $\bm{R}_A^{(0)}$ the equilibrium coordinates of atom $A$. 
To derive the frequencies and the vibronic couplings to the $E_g$ ($T_{2g}$) modes, we set $Q_u \ne 0$ ($Q_\zeta \ne 0$) and the others to zero. 
The vibronic coupling parameters were obtained by fitting the EOM-CCSD ${}^3T_{1g}$ term levels to the vibronic coupling within the ${}^3T_{1g}$ term, Eq. \eqref{Eq_VJT_3T1g}. 

To calculate the $^3T_{1g}$ states, we employed the ionization-potential (IP) variant of the EOM-CC approach \cite{Stanton1994, Pieniazek2008}, in which one electron is removed from a $t_{2g}^3$ reference state ($S=3/2$). 
To account for the contribution from the crystal environment, we replaced all the remaining 
ions in a supercell composed of $3 \times 3 \times 3$ conventional unit cells with point charges (Fig. \ref{Fig_cluster}). 
As point charges, we adopted Mulliken charges of W: $+0.99$, O: $-0.48$, and Cs: $+0.99$ from the CCSD calculations for the bare Cs$_8$WCl$_6$ cluster.
One should note that, in the present case, the presence or absence of the point charges has little influence on the electronic states, as demonstrated for the $5d^1$ double perovskite \cite{Matsuzaki2026}.

\subsection{Vibronic states}
\label{Sec_vibronic_states}
We calculated the vibronic states by numerically diagonalizing the model vibronic Hamiltonian. 
Combining Eqs. \eqref{Eq_Psi} and \eqref{Eq_chi_n}, the basis set for the vibronic states is the products of the electronic energy eigenstates and the vibrational states. 
We truncated the vibronic basis set to satisfy 
\begin{align}
&n_u + n_v + n_\xi + n_\eta + n_\zeta \le 7,
\nonumber\\
&n_u + n_v \le 5, \quad n_\xi + n_\eta + n_\zeta \le 4.
\end{align}
Furthermore, we separately calculated the vibronic eigenstates from different term states with a large energy gap, as in Ref. \cite{Iwahara2023RIXS}.
We split the electronic term basis into five groups: 
$\{{}^3T_{1g}\}$, $\{{}^1T_{2g}, {}^1E_g\}$ from the $t_{2g}^2$ electron configurations, 
$\{{}^3T_{2g}\}$, $\{{}^3T_{1g}\}$, and $\{{}^1T_{2g}, {}^1T_{1g}\}$ from the $t_{2g}^1e_g^1$ configurations.

\subsection{RIXS}
\label{Sec_RIXS}
We simulated the W $L_3$ edge RIXS spectra. 
We expressed the RIXS cross-section by using the Kramers-Heisenberg formula \cite{RIXS, Sakurai1967}: 
\begin{align}
 I_{fi} &\propto 
 \rho_{i}
 \left| 
 \sum_n
 \frac{(\bm{e}^{(f)}\cdot \langle f|\hat{\bm{p}} |n\rangle )(\bm{e}^{(i)}\cdot \langle n|\hat{\bm{p}}|i\rangle)}{\hslash \omega_i + E_i - E_n + i\frac{\Gamma}{2}} \right|^2
 \nonumber\\
 &\times
 \delta(\hslash \omega_i + E_i  -\hslash \omega_f - E_f).
\end{align}
Here, the initial ($|i\rangle$) and final ($|f\rangle$) states are vibronic states of the $5d^2$ octahedron,
$\bm{e}^{(i/f)}$ are the polarization vectors of the incident and scattered photons, 
$\hat{\bm{p}}$ the electronic momentum operator, 
$\hslash \omega_i$ the incident photon energy, 
$E_i$ the initial vibronic level, 
$E_n$ the core-hole state, 
$\Gamma$ is the lifetime of the core-hole state, 
$\rho_i \propto \exp(-E_i/k_BT)$ is the Boltzmann factor, and $k_B$ and $T$ are the Boltzmann factor and temperature, respectively. 
For the $L_3$ edge RIXS spectra, the sum $\sum_n$ is over the product states of the $5d^3$ multiplet and $2p_{j=3/2}$ core hole states.
The lifetime $\Gamma$ is about 5.5 eV for $5d$ elements \cite{Clancy2012}.

We included the covalency effect between the W $5d$ and Cl $3p$ orbitals into the transition moments. 
In the momentum operators, we described the covalency by introducing reduction factors $\eta_\Gamma$:
\begin{align}
 \hat{p}_\alpha &= \sum_{\Gamma = e_g, t_{2g}} \sum_{\gamma\in \Gamma} \sum_{\gamma_p \in 2p} \sum_{\sigma}
 \eta_{\Gamma} \langle \Gamma\gamma |\hat{p}_{\alpha} |2p\gamma_p\rangle  \hat{d}_{\Gamma\gamma\sigma}^\dagger \hat{c}_{\gamma_p\sigma}.
\end{align}
We set $\eta_{e_g}/\eta_{t_{2g}} = 0.82$ based on the {\it ab initio} calculations described below.

For the cross-section calculations, we introduced an additional approximation. 
For the scatterings to the $(e_g)^1(t_{2g})^1$ and $(e_g)^2(t_{2g})^0$ configurations, we approximated the $5d^3$ part of the intermediate energies $E_n$ with typical energies for $(e_g)^1(t_{2g})^2$ and $(e_g)^2(t_{2g})^1$ configurations, respectively, because their contributions are dominant for the final states. 
With the approximation, the cross-section reduces to 
\begin{align}
 I_{fi} &\propto 
 \rho_{i}
 |G_f|^2
 \left| 
 \sum_n
(\bm{e}^{(f)}\cdot \langle f|\hat{\bm{p}} |n\rangle )(\bm{e}^{(i)}\cdot \langle n|\hat{\bm{p}}|i\rangle)
 \right|^2
 \nonumber\\
 &\times
 \delta(\hslash \omega_i + E_i -\hslash \omega_f - E_f),
 \label{Eq_I_RIXS_FC}
\end{align}
where $G_f$ is the approximated core-hole propagator:
\begin{align}
 |G_f| &= 
 \begin{cases}
  |\Gamma/2|^{-1}, & \text{for the $(t_{2g})^2$ peaks}\\
  |10Dq + i \Gamma/2|^{-1}. & \text{for the $(t_{2g})^1(e_g)^1$ peaks}
 \end{cases}
\end{align}
We did not calculate the core-hole states by using the EOM-CC method because the further details of the multiplet structure are smaller than $\Gamma$ and ligand-field splitting, and hence, do not give a considerable effect on the scattering amplitude.
For the linewidth of the RIXS spectra, we convoluted the cross-section with the Gaussian function, $e^{-x^2/2\sigma^2}$.

The scattering geometry of our calculations of the RIXS spectra is as follows. 
The wave vectors of the incident ($\bm{k}_i$) and scattered ($\bm{k}_f$) photons are perpendicular to each other, and $\bm{k}_f - \bm{k}_i \parallel [111]$.
The incident photons are $\pi$-polarized, and we did not distinguish the polarizations of the scattered photons.

\subsection{Effective magnetic moment}
\label{Sec_chi}
We simulate the effective magnetic moment for a single $5d^2$ site. 
Adding Zeeman interaction,
\begin{align}
 \hat{H}_\text{Zee} &= -\hat{\bm{\mu}} \cdot \bm{H},
 \\
 \hat{\bm{\mu}} &= -\mu_B ( \hat{\bm{L}} + g_e \hat{\bm{S}} ),
 \label{Eq_mu}
\end{align}
to our electronic or vibronic Hamiltonian, we calculate the energy levels under an applied magnetic field. 
Here, $\hat{\bm{L}}$ and $\hat{\bm{S}}$ are the orbital angular momentum and spin angular momentum operators acting on the $d^2$ configuration space, respectively, $\mu_B$ the Bohr magneton, $g_e$ the $g$-factor of electron, $\hat{\bm{\mu}}$ the magnetic moment operator, and $\bm{H}$ the applied magnetic field. 
In this work, we applied the magnetic field along the $c$ axis.
With the obtained Zeeman levels, we calculated the Helmholtz free energy $F$, and then obtained the magnetic susceptibility: 
\begin{align}
 \chi_\text{mag} = \left.\frac{\partial^2 F}{\partial H^2}\right|_{\bm{H}=\bm{0}}.
\end{align}
We numerically calculated the second-order differentiations of $F$. 
The effective magnetic moment $M_\text{eff}$ is defined as \cite{Kotani1949, Kotani1960},
\begin{align}
M_\text{eff} &= \sqrt{3k_B T \chi_\text{mag}}. 
\label{Eq_Meff}
\end{align}

For the calculations of $M_\text{eff}$ with the vibronic model, we projected the magnetic moment operator into the vibronic states from the $^3T_{1g}$ electronic term states, and ignored the Van Vleck contribution from the excited term states.

\section{Results and discussion}
\label{Sec_result}

\subsection{Electronic structure}
\label{Sec_electronic}

\begin{table}[tb]
\caption{The interaction parameters for the electronic and vibronic models.
The units for the electronic parameters are in eV, the units for the frequencies are in meV, and the vibronic coupling parameters are dimensionless. 
See for spin-orbit coupling Table \ref{Table_SOC}.}
\label{Table_vibronic}
\begin{ruledtabular}
\begin{tabular}{lllll}
\multicolumn{2}{c}{Electronic (eV)} & \multicolumn{3}{c}{Vibronic}\\
& & & $E_g$ & $T_{2g}$ \\
\hline
$10Dq$    &  3.32 & $\omega_\Gamma$ & 38.8 & 19.7 \\ 
$B$       & 0.045 & $g_\Gamma$      & $-0.83$ & 0.42 \\
$C$       & 0.193 & $w_\Gamma$      & $-0.12$ & 0.30 \\
\end{tabular}
\end{ruledtabular}
\end{table}

\begin{table*}[htbp]
\caption{Spin-orbit coupling parameters taking account of the covalency effects, $\lambda_{{}^{[S]}\Gamma\text{-}{}^{[S']}\Gamma'}$ (eV). We build the spin-orbit Hamiltonian by substituting the parameters into Eq. (\ref{Eq_HSO_cov}).}
\label{Table_SOC}
\begin{ruledtabular}
\begin{tabular}{cc|ccccccccccc}
 & & $(t_{2g})^2$ & & & & $(t_{2g})^1(e_g)^1$ & & & & $(e_g)^2$ 
\\
 & & $^3$\text{T}$_{1\text{g}}$ & $^1$\text{T}$_{2\text{g}}$ & $^1$\text{E}$_{\text{g}}$ & $^1$\text{A}$_{1\text{g}}$
 & $^3$\text{T}$_{2\text{g}}$ & $^3$\text{T}$_{1\text{g}}$ & $^1$\text{T}$_{2\text{g}}$ & $^1$\text{T}$_{1\text{g}}$
 & $^3$\text{A}$_{2\text{g}}$ & $^1$\text{E}$_{\text{g}}$ & $^1$\text{A}$_{1\text{g}}$ \\
\hline
$(t_{2g})^2$        & $^3$\text{T}$_{1\text{g}}$ & 0.257 &  &  &  &  &  &  &  &  &  &  \\
                    & $^1$\text{T}$_{2\text{g}}$ & 0.266 & 0     &  &  &  &  &  &  &  &  &  \\
                    & $^1$\text{E}$_{\text{g}}$  & 0.253 & 0     & 0     &  &  &  &  &  &  &  &  \\
                    & $^1$\text{A}$_{1\text{g}}$ & 0.259 & 0     & 0     & 0     &  &  &  &  &  &  &  \\
$(t_{2g})^1(e_g)^1$ & $^3$\text{T}$_{2\text{g}}$ & 0.250 & 0.248 & 0.272 & 0     & 0.267 &  &  &  &  &  &  \\
                    & $^3$\text{T}$_{1\text{g}}$ & 0.250 & 0.237 & 0.243 & 0.245 & 0.274 & 0.226 &  &  &  &  &  \\
                    & $^1$\text{T}$_{2\text{g}}$ & 0.241 & 0     & 0     & 0     & 0.242 & 0.232 & 0     &  &  &  &  \\
                    & $^1$\text{T}$_{1\text{g}}$ & 0.246 & 0     & 0     & 0     & 0.263 & 0.268 & 0     & 0     &  &  &  \\
$(e_g)^2$           & $^3$\text{A}$_{2\text{g}}$ & 0     & 0.327 & 0     & 0     & 0.256 & 0     & 0.246 & 0     & 0     &  &  \\
                    & $^1$\text{E}$_{\text{g}}$  & 0.223 & 0     & 0     & 0     & 0.249 & 0.242 & 0     & 0     & 0     & 0 &  \\
                    & $^1$\text{A}$_{1\text{g}}$ & 0.785 & 0     & 0     & 0     & 0     & 0.191 & 0     & 0     & 0     & 0 & 0 \\
\end{tabular}
\end{ruledtabular}
\end{table*}

We obtained the term levels by EOM-CCSD calculations. 
Figure \ref{Fig_TS}(a) compares the calculated term levels with the Tanabe-Sugano diagram.
We extracted the ligand-field ($10Dq$) and Racah's parameters ($B, C$) by fitting the electronic model to the low-lying EOM-CCSD term levels ($^3T_{1g}$, $^1T_{2g}$, $^1A_{1g}$, $^3T_{2g}$, $^3T_{1g}$):
$10Dq = 3.32$ eV, $B = 0.045$ eV, and $C = 0.193$ eV (Table \ref{Table_vibronic}).
The present $10Dq$ and the previous DFT value of about 3 eV \cite{Pradhan2024} with the Perdew-Burke-Ernzerhof exchange correlation functional \cite{PBEsol} are within 11\% of each other. 
The ratio $C/B = 4.29$ is close to the Hartree-Fock value for isolated W$^{4+}$ ($C/B = 4.12$ \cite{Ma2014}).
With the present $B$ and $C$, the Hund's rule coupling $J_H = 3B + C = 0.329$ eV.

Using the EOM-CCSD term states obtained, we calculated the Breit-Pauli spin-orbit Hamiltonian matrix. 
Then, by fitting the {\it ab initio} spin-orbit Hamiltonian matrix to the model \eqref{Eq_HSO_cov}, we determined the spin-orbit coupling constants. 
Table \ref{Table_SOC} lists all the spin-orbit coupling parameters.
The spin-orbit coupling parameter for the ${}^3T_{1g}$ ground term states is similar to a DFT estimate of atomic $\lambda=0.27$ eV \cite{Li2025}. 
The $5d$-$3p$ covalency reduces the spin-orbit coupling parameters, and the extent of the reduction depends on the terms.

Substituting all the derived interaction parameters into the electronic model comprising Hund's, ligand-field, and spin-orbit coupling, we calculated the multiplet energy levels [Fig. \ref{Fig_TS}(b)].
The spin-orbit coupling splits the ground $^3T_{1g}$ term states into the multiplet states characterized by effective $J_\text{eff}= 2, 1, 0$  in the increasing order of energy. 
The low-symmetric environment lifts the $J_\text{eff}=2$ multiplet states into the $E_g$ and the $T_{2g}$ levels.
The $E_g$ state is lower in energy than the $T_{2g}$ state. 
The splitting occurs due to the nonspherical Hund's coupling and the spin-orbit coupling between the $J_\text{eff}=2$ and the excited $t_{2g}^1e_g^1$ term states \cite{Voleti2020, Takayama2021, Pradhan2024}.
The obtained $E_g$-$T_{2g}$ gap is about 14.7 meV [Fig. \ref{Fig_TS}(c)].

\begin{figure}[tb]
\includegraphics[width=\linewidth, bb=0 0 504 310]{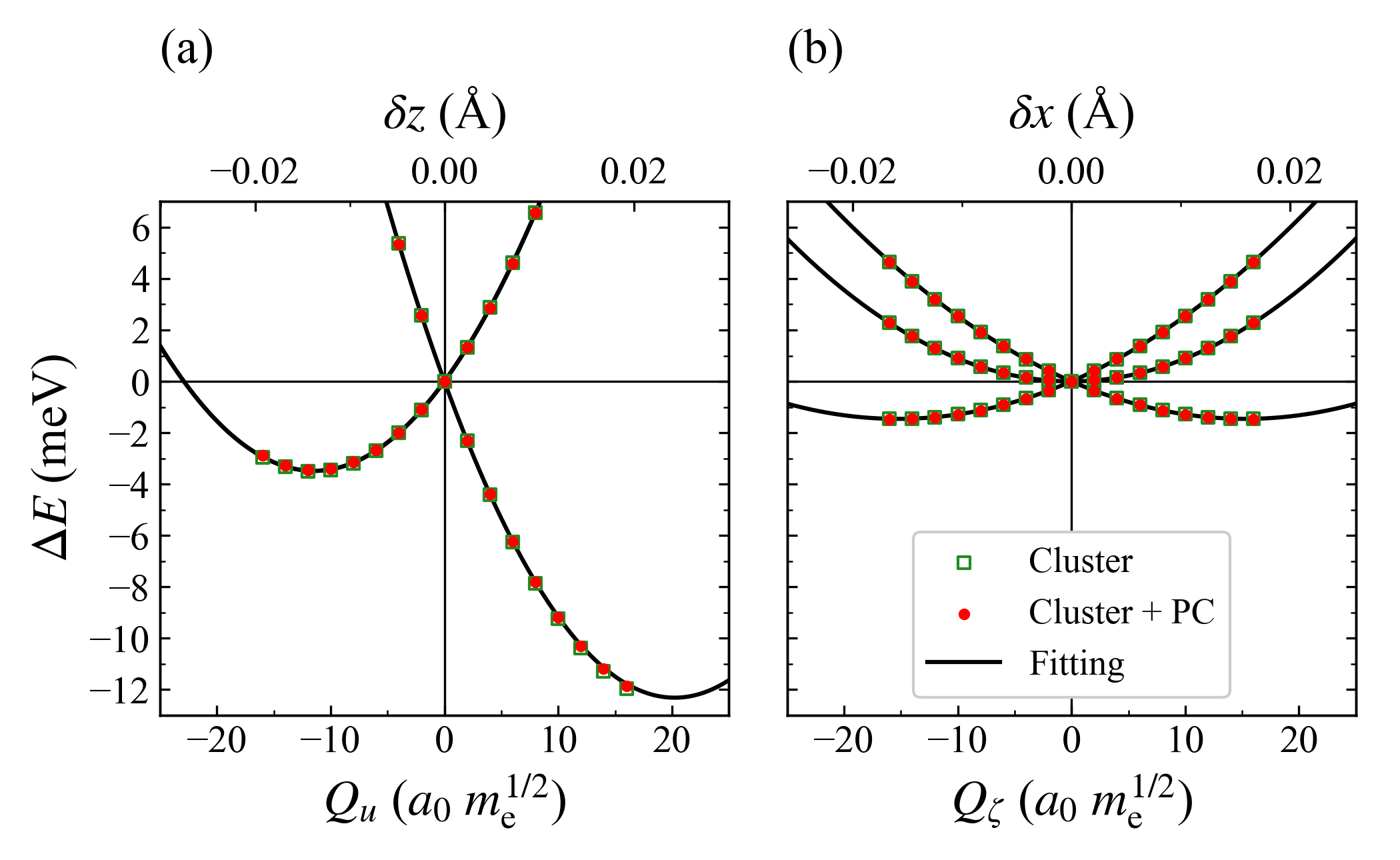}
\caption{The $^3T_{1g}$ APESs with respect to (a) the $E_gu$ and (b) the $T_{2g}\zeta$ deformations.
The green open squares and the red circles are the EOM-CCSD data without and with point charges, respectively. 
The black lines are the model fit to the EOM-CCSD data with point charges. 
}
\label{Fig_JT_2T2g}
\end{figure}

\begin{figure}[tb]
\includegraphics[width=\linewidth, bb=0 0 504 310]{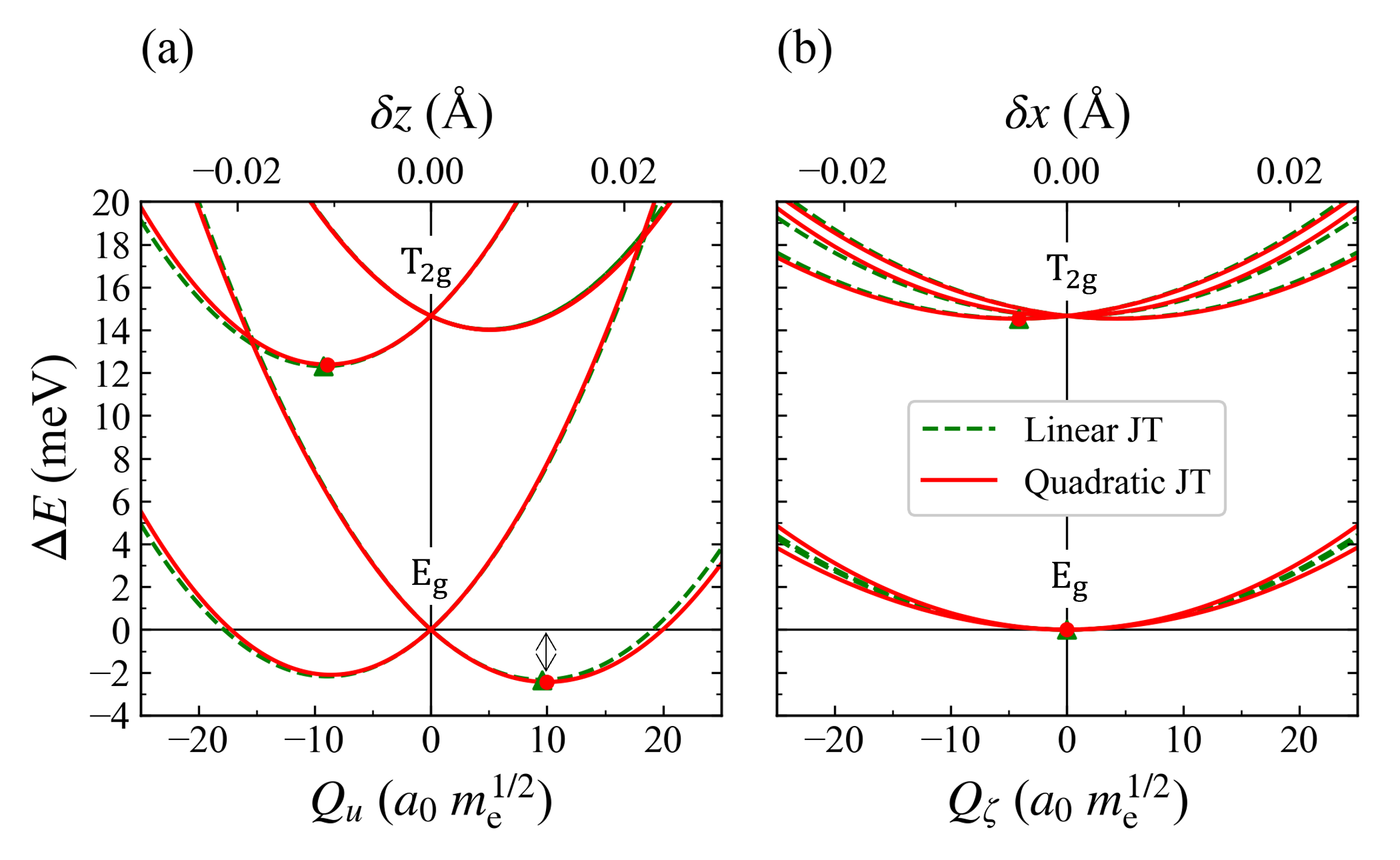}
\caption{The $J_\text{eff}=2$ APESs with respect to (a) the $E_gu$ and (b) the $T_{2g}\zeta$ deformations.
The green dashed and the red solid lines show the APESs of the models that include only linear vibronic coupling and include both linear and quadratic vibronic couplings, respectively.  
}
\label{Fig_JT_J2}
\end{figure}

\subsection{Vibronic coupling and Jahn-Teller effect}
\label{Sec_vibronic}

We determined the vibronic coupling parameters using the ${}^3T_{1g}$ EOM-CCSD adiabatic potential energy surface (APES) with respect to the JT active modes [Fig. \ref{Fig_AF}(b)]. 
Figure \ref{Fig_JT_2T2g} shows the calculated $^3T_{1g}$ APESs (red circles) and the fitted model (black lines).
Table \ref{Table_vibronic} lists the frequencies and the vibronic coupling parameters obtained by fitting the {\it ab initio} data to the model vibronic Hamiltonian \eqref{Eq_VJT_3T1g}.
The calculated data show that vibronic coupling to the $E_g$ modes is stronger than that to the $T_{2g}$ modes. 

To quantitatively examine the influence from the surrounding point charges on the $5d^2$ octahedron, we calculated the vibronic couplings and frequencies using the cluster model without surrounding point charges. 
The presence or absence of the point charges does not significantly affect the shape of the APESs, and hence, the frequencies and the vibronic couplings [see the red circles and green open squares in Fig. \ref{Fig_JT_2T2g}], as in the previous study on a $5d^1$ system \cite{Matsuzaki2026}.

Figure \ref{Fig_JT_J2} shows the low-lying APESs from the $J_\text{eff}=2$ multiplet states.  
We calculated the APES using the potential term of our DJT model Hamiltonian, $\hat{H}_\text{el} + \hat{H}_\text{DJT}$, with the {\it ab initio} interaction parameters. 
The green-dashed and red-solid lines indicate the APESs from the linear JT and quadratic JT models, respectively. 
They do not differ much in the present case. 
The static JT stabilization energies for the $E_g$ modes are 2.4 meV and 2.3 meV, respectively, in the $E_g$ and the $T_{2g}$ multiplet APESs.
The static JT stabilization energy for the $T_{2g}$ modes in the $T_{2g}$ multiplet APES is 0.1 meV. 
Overall, the vibronic coupling in the $5d^2$ Cs$_2$WCl$_6$ is a few orders of magnitude smaller than the frequencies, which indicates that the system is close to the weak JT regime.

By numerically diagonalizing the obtained DJT model, we obtained the vibronic energy levels. 
Figure \ref{Fig_TS}(c) shows the calculated levels. 
The DJT effect enhances the stabilization energy (5.0 meV) for the ground $E_g$ state, which reaches about 2 times the static JT energy. 
The DJT stabilization energy and magnitude of the intersite interactions in the compounds are slightly larger than the experimental Weiss constant of 3.8 meV \cite{Takayama2026}, suggesting the persistence of the DJT effect in the compound. 

The DJT effect varies the energy gap between the ground $E_g$ and the first excited $T_{2g}$ levels.
Vibronic coupling to the $E_g$ modes increases the gap, whereas that to the $T_{2g}$ modes decreases it.  
Using the vibronic model involving only the $E_g$ mode, the $E_g$-$T_{2g}$ energy gap is 16.2 meV. 
Including both the $E_g$ and $T_{2g}$ modes into the model, the gap becomes 15.5 meV. 
Compared with the electronic model, the vibronic couplings do not cause a noticeable change in the energy gap in Cs$_2$WCl$_6$.

With the EOM-CCSD method, we calculated the vibronic model and showed it to be in the weak-coupling regime. 
The DJT stabilization energy in the ground $E_g$ state is 5.0 meV, and it slightly widens the $E_g$-$T_{2g}$ energy gap from the electronic value of 14.7 meV to 15.5 meV. 
Due to weak vibronic coupling, JT deformations are difficult to detect experimentally.

\begin{figure*}[tb]
\includegraphics[width=\linewidth, bb=0 0 648 288]{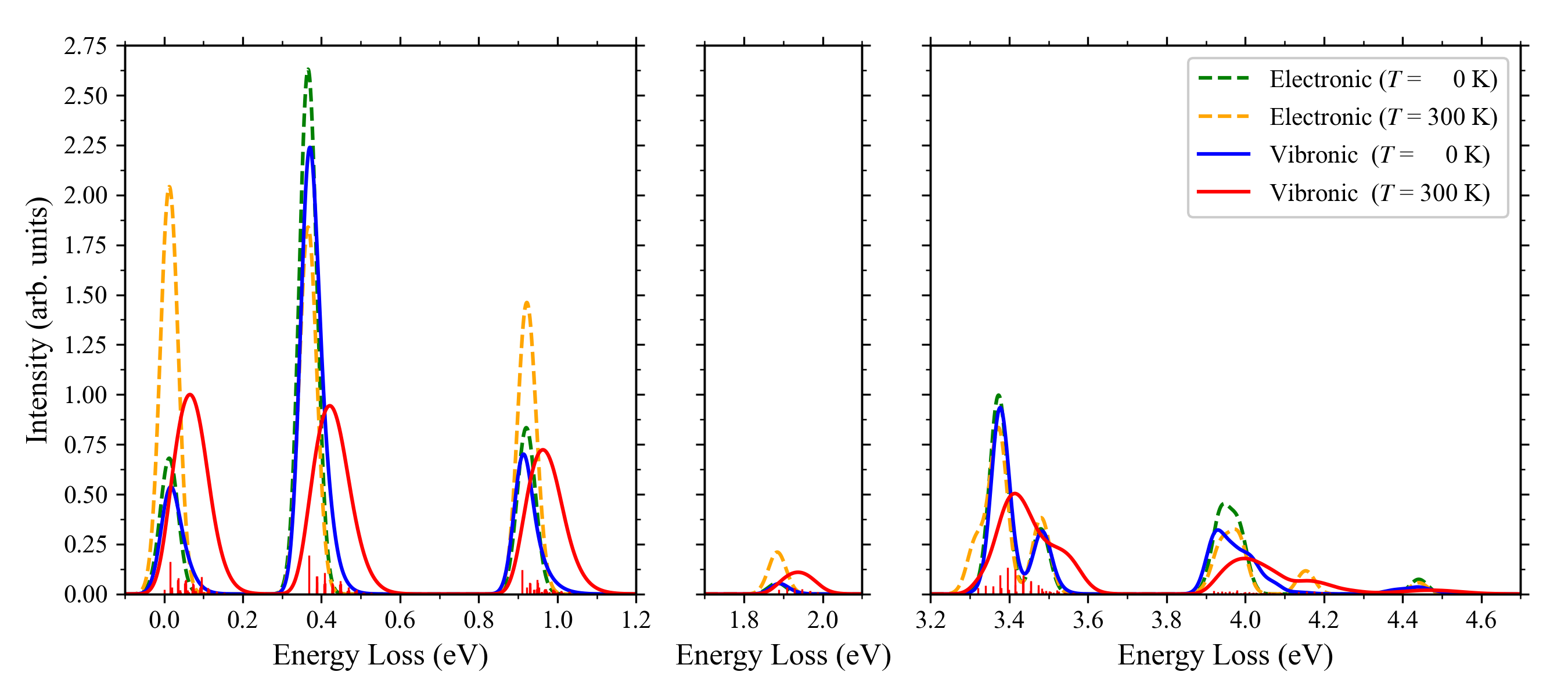}
\caption{The electronic and vibronic W $L_3$ edge RIXS spectra at 0 K and 300 K.
The green-dashed and orange-dashed lines are the electronic RIXS spectra at 0 K and 300 K, respectively.
The blue solid and red solid lines are the vibronic RIXS spectra at 0 K and 300 K, respectively.
The Gaussian broadening $\sigma = $ 0.022 eV.}
\label{Fig_RIXS}
\end{figure*}

\subsection{W $L_3$ edge RIXS spectra}
\label{Sec_RIXS}
We computed the W $L_3$ edge RIXS spectra [Fig. \ref{Fig_RIXS}]. 
The electronic RIXS spectra at 0 K exhibit symmetric peaks corresponding to transitions from the ground $E_g$ multiplet to excited levels (the green dashed lines). 
Raising the temperature to 300 K, the transitions are not only from the $E_g$ but also from the thermally populated first-excited $T_{2g}$ multiplet states (the orange dashed lines). 
The scattering from the $T_{2g}$ level enhances the heights of some peaks, such as the elastic peak and the one from ${}^1E_g/{}^1T_{2g}$ term states.

The shape and the temperature dependence of the vibronic W $L_3$ edge RIXS spectra (the blue solid and red solid lines) are overall similar to the electronic ones. 
The vibronic effect manifests as a change in the peak shape in the present case. 
Vibronic coupling makes the peaks asymmetric compared to the electronic ones. 
The asymmetry arises from the multitude of transitions from the ground $E_g$ to the excited vibronic states, as in the other cases \cite{Iwahara2023RIXS, Iwahara2023RIXS2, Frontini2024, Warzanowski2024, Iwahara2025}. 

The simulated spectra share common features with recently measured W $L_3$ edge RIXS spectra of $A_2$WCl$_6$ ($A=$ Rb, Cs) \cite{Takayama2026}.
The experimental RIXS spectra at 8 K have sharp peaks at the term energies.
Moreover, some of the peaks for $J_\text{eff}=1, 0$ ($\approx 0.4$ eV), ${}^1E_g$, ${}^1T_{2g}$ ($\approx 1$ eV) are asymmetric and the one for $^1A_{1g}$ remains symmetric. 
These features are consistent with the experimental W $L_3$ edge RIXS spectra \cite{Takayama2026}.

Our calculated spectra underestimate the peak corresponding to the transition from the ground $E_g$ state to the first excited $T_{2g}$ state. 
The experimental elastic peak at low temperature \cite{Takayama2026} is as strong as our room-temperature peak.
The discrepancy between the theoretical and experimental spectra may arise from intersite interactions between the W centers in Cs$_2$WCl$_6$, which is not considered in the present work. 

\begin{table*}[htbp]
\caption{Reduction factor for the orbital angular momenta, $k_{{}^{[S]}\Gamma\text{-}{}^{[S']}\Gamma'}$, due to $5d$-$3p$ covalency.  
We can build the orbital angular momenta matrices by using the parameters and Eq. (\ref{Eq_L_cov}).}
\label{Table_L}
\begin{ruledtabular}
\begin{tabular}{cc|ccccccccccc}
 & & $(t_{2g})^2$ & & & & $(t_{2g})^1(e_g)^1$ & & & & $(e_g)^2$ 
\\
 & & $^3$\text{T}$_{1\text{g}}$ & $^1$\text{T}$_{2\text{g}}$ & $^1$\text{E}$_{\text{g}}$ & $^1$\text{A}$_{1\text{g}}$
 & $^3$\text{T}$_{2\text{g}}$ & $^3$\text{T}$_{1\text{g}}$ & $^1$\text{T}$_{2\text{g}}$ & $^1$\text{T}$_{1\text{g}}$
 & $^3$\text{A}$_{2\text{g}}$ & $^1$\text{E}$_{\text{g}}$ & $^1$\text{A}$_{1\text{g}}$ \\
\hline
$(t_{2g})^2$        &$^3$\text{T}$_{1\text{g}}$ & 0.800 &  &  &  &  &  &  &  &  &  &  \\
                    &$^1$\text{T}$_{2\text{g}}$ & 0     & 0.979 &  &  &  &  &  &  &  &  &  \\
                    &$^1$\text{E}$_{\text{g}}$  & 0     & 0.877 & 0     &  &  &  &  &  &  &  &  \\
                    &$^1$\text{A}$_{1\text{g}}$ & 0     & 0     & 0     & 0     &  &  &  &  &  &  &  \\
$(t_{2g})^1(e_g)^1$ &$^3$\text{T}$_{2\text{g}}$ & 0.359 & 0     & 0     & 0     & 0.874 &  &  &  &  &  &  \\
                    &$^3$\text{T}$_{1\text{g}}$ & 0.251 & 0     & 0     & 0     & 0.989 & 0.613 &  &  &  &  &  \\
                    &$^1$\text{T}$_{2\text{g}}$ & 0     & 0.360 & 3.354 & 0     & 0     & 0     & 0.668 &  &  &  &  \\
                    &$^1$\text{T}$_{1\text{g}}$ & 0     & 0.308 & 0.341 & 0.291 & 0     & 0     & 0.824 & 0.876 &  &  &  \\
$(e_g)^2$           &$^3$\text{A}$_{2\text{g}}$ & 0     & 0     & 0     & 0     & 0.356 & 0     & 0     & 0     & 0     &  &  \\
                    &$^1$\text{E}$_{\text{g}}$  & 0     & 0.438 & 0     & 0     & 0     & 0     & 0.345 & 0.305 & 0     & 0 &  \\
                    &$^1$\text{A}$_{1\text{g}}$ & 0     & 0     & 0     & 0     & 0     & 0     & 0     & 0.154 & 0     & 0 & 0 \\
\end{tabular}
\end{ruledtabular}
\end{table*}

\begin{figure}[tb]
\includegraphics[width=\linewidth, bb=0 0 324 252]{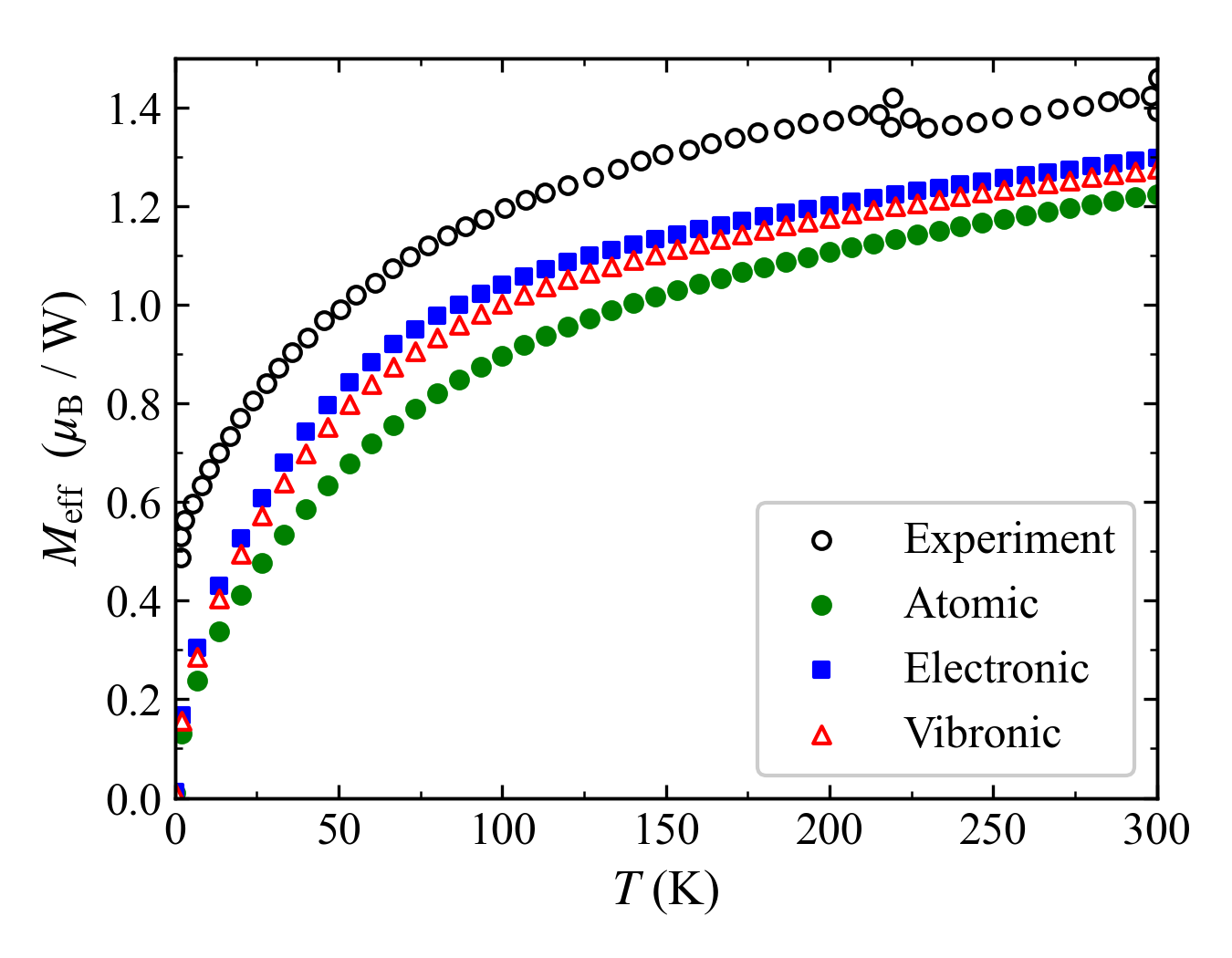}
\caption{Temperature dependence of $M_\text{eff}$ per W with applied magnetic field of $B=0.05$ T. 
The black points are the experimental data. 
We subtracted the diamagnetic correction [(molar mass/2) $\times 10^{-6}$] from the data in Ref. \cite{Morgan2023} as was done in Ref. \cite{Li2025}.
The blue squares and the green circles are, respectively, the electronic $M_\text{eff}$ with {\it ab initio} and an atomic spin-orbit couplings, respectively, and the red open triangles are the vibronic $M_\text{eff}$. 
}
\label{Fig_Meff}
\end{figure}

\subsection{Effective magnetic moment}
\label{Sec_Meff}
To obtain $M_\text{eff}$, we calculated the magnetic moment operators \eqref{Eq_mu} with the EOM-CCSD method. 
The orbital angular momentum operators differ from the atomic ones due to the W $5d$-Cl $3p$ hybridization as discussed in Sec. \ref{Sec_chi}. 
Table \ref{Table_L} lists the calculated reduction factors for orbital angular momenta. 
As in the spin-orbit coupling, the values of the reduction factors depend on the electronic term states. 

With the calculated orbital angular momenta, we simulated the temperature-dependent effective magnetic moment \eqref{Eq_Meff}.
Figure \ref{Fig_Meff} shows the calculated $M_\text{eff}$ with respect to $T$.
We applied the magnetic field along the $c$ axis. 
The blue-filled circles represent the calculated electronic $M_\text{eff}$, and the red triangles represent the calculated vibronic $M_\text{eff}$.
Both the electronic and vibronic $M_\text{eff}$'s are zero at 0 K, grow with temperature, and reach about 1.3 $\mu_B$.  
The temperature dependence of $M_\text{eff}$ changes around 50 K due to the thermal excitations into the $T_{2g}$ state. 
The difference between the electronic and vibronic $M_\text{eff}$'s is not significant due to weak vibronic coupling.
The slight reduction in the vibronic $M_\text{eff}$ compared with the electronic one is due to the vibronic reduction \cite{Englman1972} and the ignored Van Vleck correction from the excited-state terms.

The obtained $M_\text{eff}$'s show a close temperature dependence with the experimental data. 
In Fig. \ref{Fig_Meff}, the black circles are the experimental data \cite{Morgan2023}. 
Both the calculated and experimental $M_\text{eff}$'s bend at around 50 K due to the thermal population in the $T_{2g}$ states, supporting the validity of the calculated $E_g$-$T_{2g}$ gap of $\approx$ 15 meV. 

The calculated $M_\text{eff}$ values, however, are underestimated by about 10\% relative to the experimental data.
At high temperature, the effective magnetic moment without Van Vleck correction reaches about 1.4 $\mu_B$ \cite{Morgan2023, Takayama2026}.
On the theoretical side, the reduction can occur due to insufficient $ 5d$–$3p$ hybridization. 
Improving the hybridization effect is beyond the scope of this study.

The present calculations bring further qualitative insight into the term-dependent covalency effect in $M_\text{eff}$.
We compare the $M_\text{eff}$ with the term-dependent $\hat{\bm{\mu}}$ and that with an atomic model. 
The green circles in Fig. \ref{Fig_Meff} show the electronic results with an atomic spin-orbit coupling and orbital angular momenta ($\lambda=0.257$ eV and $\langle L \rangle =0.800$, calculated with the ground  $^3T_{1g}$ states). 
The discrepancy between the atomic model and the experimental data is larger than that between the {\it ab initio} and the experimental data. 
Since $t_{2g}$-$e_g$ hybridization is important for the simulations of $M_\text{eff}$ \cite{Stamokostas2018, Li2025}, the orbital dependence of the covalency effect influences the temperature dependence of $M_\text{eff}$.

Finally, we compare the present and published theoretical data. 
Previously estimated $E_g$-$T_{2g}$ gaps are 15 meV \cite{Pradhan2024} and 30 meV \cite{Li2025}. 
The current energy gap of 15.5 meV is close to them, though caution is warranted, as these works relied on simplified electronic models.
In the former work \cite{Pradhan2024}, Kanamori's model \cite{Kanamori1963} with atomic $\lambda$ and $J_H$ from the data of K$_2$OsCl$_6$ \cite{Warzanowski2023} was employed. 
The model ignores the covalency effects and the influence of the $e_g$ orbitals.  
Moreover, to estimate the $E_g$-$T_{2g}$ gap, the ligand-field splitting is set to only 1.7 eV, which is about half of the present {\it ab initio} data and experimental RIXS data \cite{Takayama2026}.
In the latter work \cite{Li2025}, the employed model resembles our atomic model, which does not account for covalency effects.
As discussed above, $M_\text{eff}$ depends on the magnitudes of the energy gaps and the degree of covalency.  
The EOM-CC method accurately describes both effects and is therefore suitable for predicting the effective magnetic moment.

We calculated $M_\text{eff}$ of a single W site. 
The temperature dependence of $M_\text{eff}$ agrees well with the experimental data, which indicates that the predicted $E_g$-$T_{2g}$ splitting is reliable. 
Moreover, we found that intricate W $ 5d$-Cl $3p$ covalency is important for an accurate quantitative description of $M_\text{eff}$.
We stress that the prediction was achieved {\it ab initio} without fitting interaction parameters to experimental data.

\section{Conclusion}
We analyzed the vibronic, magnetic, and spectroscopic properties of an embedded $5d^2$ W ion in Cs$_2$WCl$_6$ based on the EOM-CC method. 
We derived the model Hamiltonian for the W sites, calculated the electronic and vibronic levels, and reproduced the W $L_3$ edge RIXS and effective magnetic moment. 
The agreement between the theoretical and experimental \cite{Takayama2026} RIXS spectra indicates the validity of the calculated electronic levels and the magnitude of the vibronic coupling. 
The agreement in temperature evolution between the experimental and theoretical effective magnetic moments indicates the validity of the calculated low-energy structure of the system.  
The present calculations provide a solid basis for investigating the multipolar orderings in Cs$_2$WCl$_6$. 
The same {\it ab initio} approach will apply to the analysis of the family of the $5d^2$ double perovskites. 

The calculations above and our recent work \cite{Matsuzaki2026} demonstrate the utility of the EOM-CC method for predicting the vibronic, magnetic, and spectroscopic properties of embedded metal ions in correlated insulators. 
The present EOM-CC-based approach will accelerate the understanding and prediction of complex phenomena in correlated materials.

\begin{acknowledgments}
We would like to thank T. Takayama, H. Takagi, and K. Ishii for fruitful discussions and for sharing their unpublished data, and Y. Li for useful information on effective magnetic moments.
T.M. and N.I. are grateful to the Division of Quantum and Physical Chemistry at KU Leuven for hospitality. 
The work at Chiba was partly supported by 
Grant-in-Aid for Scientific Research (Grant No. 22K03507) from the Japan Society for the Promotion of Science, 
the Nippon Sheet Glass Foundation for Materials Science and Engineering, 
and the JST CREST project led by T. Omatsu (No. JPMJCR1903).
\end{acknowledgments}

\section*{Data Availability}
The data that support the findings of this article are openly available \cite{data}.


%
\end{document}